# Tunable Rapid Electron Transport in Titanium Oxide Thin Films

Runze Li,[1,2,†] Faguang Yan,[1,2] Yongcheng Deng,[1,2] Yu Sheng[1,2]

[1]*State Key Laboratory for Superlattices and Microstructures, Institute of Semiconductors, Chinese Academy of Sciences, Beijing 100083, China*

[2] *Center of Materials Science and Optoelectronics Engineering, University of Chinese Academy of Sciences, Beijing 100049, China*

**Abstract:** Rapid electron transport in the quantum well triggers many novel physical phenomena and becomes a critical point for the high-speed electronics. Here, we found electrical properties of the titanium oxide changed from semiconducting to metallic as the degree of oxidation decreased and Schottky quantum well was formed at the interface. We take the asymmetry interface electron scattering effect into consideration when studying the electrical transport properties of the multilayer thin films. A novel physical conductivity model for the multilayer thin films was developed. We found electron would be transferred from the low-mobility semiconducting and metallic conductive channels to the high-mobility Schottky quantum well conductive channel with an in-plane applied electric field. Electron concentration and mobility of the forming 2DEG in the Schottky quantum well could be tuned thus the nano-devices exhibited non-linear voltage-current curves. The differential resistivity of the nano-devices could decrease by two orders with increasing electric field at room temperature. Weak electron localization of electrons has been experimentally observed in our nano-devices at low temperature, which further demonstrated the existence of 2DEG in the Schottky quantum well. Our work will provide us new physics about the rapid electron transport in the multilayer thin films, and bring novel functional devices for the modern microelectronic industry.

For decades, the electron transport in the thin films is one of the key issues in condensed matter physics. A variety of physical phenomena have been discovered such as superconductivity, ballistic transport and two-dimensional electron gas (2DEG) [1-3]. Extensive efforts have been made to tune these rapid electron transport effectively, which can produce novel physics phenomena and new device applications [4,5]. To investigate the rapid electron transport, heterostructures with potential well are always need to be constructed. For example, in semiconductor heterostructures, a potential well can be formed at interface and the mobility of the electrons is enhanced due to the electron spatial motions are confined in the potential well [6,7]. Another class of such structures is the metal-semiconductor (M-S) heterojunctions, in which electron transport across the junction has attracted significant interests for their fascinating optical and electrical properties [8-12]. In case of the work function of the semiconductor is higher, charge transfer from metal into semiconductor will take place and Schottky quantum well can be formed at the M-S interface. Little attention has been concerned to the in-plane rapid electron transport of this potential well. Generally, parallel conductance model is used to describe the in-plane electron transport of the multilayer thin films, in which conductivity of each layer is fixed and these conductive channels do not affect each other. However, for the thin films in our study, we found parallel conductance model is not applicable. Electrically tunable electron concentration redistribution among different conductive channels plays an important role for the in-plane electron transport of this Schottky quantum well.

Titanium thin films were grown at room temperature on $SiO_2$ terminated single-crystal Si substrate by magnetron sputtering. The clean samples were subsequently placed in the air, allowing the surface of the titanium films to be oxidized naturally. The XPS spectrum of titanium $2p_{3/2}$ and $2p_{1/2}$ for the 10 nm thick oxidized film and peak fitting were shown in Fig. 1(a). The $Ti^{4+}$ peak intensity at 458 eV was strongest, indicating that titanium on the film surface was mainly oxidized to $TiO_2$. The XPS spectrum could be deconvoluted into four peaks, corresponding to the titanium atoms with four different valence states of 4+, 3+, 2+, 0, which suggested that parts of the interior titanium atoms were still not oxidized. Energy dispersive X-ray spectroscopy (EDS) dot-scans were used to examine the elements distribution inside the film. The extracted atomic fraction of titanium and oxygen in the out-of-plane direction was shown in Fig. 1(b). Atomic fraction of titanium atoms decreased with the testing region getting closer to the film surface whereas oxygen atoms increased simultaneously. Therefore, depending on the degree of oxidation, in the out-of-plane direction, the transition from the $TiO_2$ on the film surface to the $Ti_2O$ within the film was observed in our samples. Fig. 1(c) shows high-resolution TEM images of the naturally oxidized titanium thin film. The room-temperature-deposited titanium film was proved to be polycrystalline and the formed $TiO_2$ near the film surface was partially crystallized in the form of islands. It has been revealed that the polycrystalline titanium oxide $TiO_\delta$ ($1 < \delta \leq 2$) thin films could exhibit p and n type semiconducting properties at high and low degree of oxidation respectively and $Ti_2O$ presented metallic electrical properties [13-15]. As will be proved below, with the degree of oxidation decreased, electrical properties of p, n type $TiO_\delta$ semiconductor and metallic $Ti_2O$ were presented in the films sequentially. Since the work function of n type $TiO_\delta$ semiconductor was higher than the metallic $Ti_2O$ [16], a Schottky quantum well was formed at the n-type $TiO_\delta/Ti_2O$ interface. This structure provided us with suitable devices to investigate the in-plane rapid electron transport properties of the Schottky quantum well.

Fig. 1(d) shows a schematic diagram of the Hall bar device and the six-probe technique was used for the longitude and Hall voltage measurements. Constant pulse currents $I_X$ with a duration of 0.3 s were applied across the device and the longitude and Hall voltages were measured by two voltmeters. Fig. 1(e) shows the voltage-current *(VI)* curves for devices with various thicknesses, where $V_X$ is the longitude voltage. When $I_X$ = 0.08 mA, differential resistivity of the device $\rho_X$ decreased from 320 to 75 $\mu\Omega$ cm as the thickness increased from 8 to 25 nm. This is consistent with other studies that as the film thickness increases, the differential resistivity decreases and eventually approaches the value of bulk titanium (42 $\mu\Omega$ cm) [17]. However, the behavior was obviously different when $I_X$ = 2 mA. As shown in the Fig. 1(f), two distinctive characteristics could be observed. Firstly, the value of the differential resistivity was very small, only 15 $\mu\Omega$ cm for the 8 nm thick film-based device, which demonstrated the existence of the rapid electron transport. Secondly, as the device thickness decreased, the reduction degree of the differential resistivity induced by the increasing current intensity was enhanced, indicating that this rapid electron transport was induced by interface rather than an effect of the bulk phase. As previously discussed, in our nano-device structure, the in-plane rapid electron transport only could be achieved by the 2DEG in the Schottky quantum well. For the 8 nm thick film-based device, as the longitude voltages increased from 0.19 to 0.5 V, differential resistivity decreased from 320 to 15 μΩ cm, indicating that electrical properties of the 2DEG formed in this Schottky quantum well could be controlled by the in-plane electric field.

The formed electrically tunable 2DEG in the Schottky quantum well was attributed to the asymmetric interface electron scattering effect in the multilayer nano-devices. Many reported studies have shown that the electron scattering at surfaces and interfaces is important for the conductive behavior of the nanoscale thin films [18,19]. For the homogeneous single layer films, all electrons are reflected back to their original layer after the surface scattering. However, for the multilayer films, electrons can be not only reflected back to their original layer but also transmitted to the adjacent layer after the interface scattering. This process can be described by the electron reflection coefficient $P_R$ and transmission coefficient $P_T$ respectively. Reflection and transmission electron current density can be obtained by the equations: $N_R = P_R\ n\ \mu\ E_{in}$ and $N_T = P_T\ n\ \mu\ E_{in}$, where $n$, $\mu$, $E_{in}$ are electron concentration, mobility and in-plane electric field intensity respectively. As shown in Fig. 2(a), for a bilayer film, if the electron concentration and mobility between the adjacent A and B layers differ greatly, the inflow electron current density does not equal to the outflow electron current density for each layer on applying $E_{in}$ ($N_{A-B} \neq N_{B-A}$). Electron concentration will be redistributed between A and B layers until the inflow and outflow electron current density reach the same value ($N_{A-B} = N_{B-A}$). If this electron concentration redistribution does not change with $E_{in}$, the conductive behavior of the bilayer film still can be described by the parallel conductance model. However, for the titanium oxide film-based nano-devices in our study, electron concentration redistribution among different conductive channels was subject to $E_{in}$.

A simplified physical conductivity model was developed to qualitatively explain our experiment results. Based on the analysis results of Fig. 1, the conductive region in our nano-devices could be divided into four channels along the out-of-plane direction, which were p, n type $TiO_\delta$ semiconductor, Schottky quantum well and metallic $Ti_2O$ channel respectively, as shown in Fig. 2(b). To simplify the model, we assumed that regions inside each channel were homogeneous. We mainly focus on the electron concentration redistribution among different conductive channels and its influence on the 2DEG quasi-ballistic transport in the Schottky quantum well. Electrons being scattered into the potential well were mostly derived from the adjacent n type $TiO_\delta$ semiconductor and metallic $Ti_2O$ conductive channels. On the metallic $Ti_2O$ side, the energy band was upward near the interface due to the electron migration when the Schottky quantum well was formed. As shown in Fig. 2(c) and Fig. 2(d), due to the existence of the finite potential barrier $U_0$, the transmission coefficient $P_4(E_X)$ was affected by the amplitude of the electron drift velocity in the metallic conductive channel. A larger $E_X$ leads to a higher electron drift velocity and hence to a larger $P_4(E_X)$. The inflow electron current density for the Schottky quantum well $N_{in}$ could be obtained by the equation $N_{in}= (P_2\mu_2 n_2 + P_4(E_X)\mu_4 n_4) E_X$, where $P_2$, $\mu_2$, $n_2$, $P_4(E_X)$, $\mu_4$, $n_4$ were electron transmission coefficient, mobility and concentration of the n type semiconductor and metallic conductive channels respectively and all of these parameters did not change with $E_X$ except $P_4(E_X)$. For the Schottky quantum well, electron transport parameters including transmission coefficient $P_3$, mobility $\mu_3$ depended on the value of the electron concentration in the potential well $n_3$. A larger $n_3$ leads to a deeper potential well and hence to a higher $\mu_3(n_3)$ and a lower value of $P_3(n_3)$ [20]. Outflow electron current density for the Schottky quantum well could be obtained by the equation $N_{out} = P_3(n_3) \mu_3(n_3) n_3(E_X, I_X) E_X$. In the stable state, $N_{in}$ must equal to $N_{out}$ and current intensity flowing through the entire nano-device $I_X = q (\mu_1 n_1 S_1 + \mu_2 n_2 S_2 + \mu_3 n_3 S_3 + \mu_4 n_4 S_4) E_X$, where $S_1$ to $S_4$ were the cross-sectional area of each corresponding conductive channels. With increasing $E_X$, $N_{in}$ increased firstly and then more electrons were distributed in the Schottky quantum well to make $N_{out}$ increase until the steady state condition ($N_{in} = N_{out}$) was achieved. Considering that value of $\mu_3$ was larger than $\mu_2$ and $\mu_4$, the differential resistivity of the entire nano-device would decrease because electrons were transferred from the n type $TiO_\delta$ semiconductor and metallic $Ti_2O$ to the Schottky quantum well conductive channel.

In the steady state, the value of $N_{in}$ and $N_{out}$ only depended on $V_X$. However, due to $P_3$ decreased with the increasing $n_3$, the same value of $N_{out}$ can correspond to two different electron concentration distribution states. For the situation with relatively large $n_3$ and small $P_3$, the resistivity of the device is lower than the other situation (small $n_3$ and large $P_3$) with the same $V_X$. The above inference was supported by the $VI$ curves obtained at low temperature. As shown in Fig. 3(a), when the temperature dropped below 12 K, intensity of $V_X$ increased firstly and then decreased with the increasing $I_X$. Two different current states could exist stably under the same $V_X$ when it was in the particular range. The constant current pulses were supplied by a Keithley 6221 current source and the changing process of output voltage signals to reach the steady states were collected by an oscilloscope. As shown in Fig. 3(b), for the *VI* curve with $T$=4 K, when the applied constant pulse was 50 μA, the output voltage increased linearly from 0 to 2.08 V and the Schottky quantum well possessed relatively small $n_3$ and large $P_3$ in this case. When the applied constant pulse was 2000 μA, as shown in Fig. 3(c), the output voltage current increased firstly to let electrons be scattered into Schottky quantum well and then decreased to satisfy the steady state condition ($N_{in} = N_{out}$). The Schottky quantum well possessed relatively large $n_3$ and small $P_3$ in this case.

The electrical properties of 2DEG in the Schottky quantum well were further investigated from Hall-effect measurements. The external magnetic field along z direction $B_Z$ was applied and swept between $\pm 6$ T. The variation of Hall and longitude voltages with the increasing $I_X$ at low temperature ($T$=3 K) were shown in Fig. 3(d). As the $I_X$ increased from 0 to 0.1 mA, both the amplitude of Hall voltage $V_{Hall}$ and $V_X$ increased linearly. This behavior was consistent with the Hall effect in most conventional semiconductor materials and the negative value of $V_{Hall}$ indicated the existence of hole carriers. When the $I_X$ increased from 0.1 to 0.2 mA, increasing rate of $V_X$ became slow and *VI* curve no longer kept linear. Amplitude of the negative $V_{Hall}$ decreased and sign of the $V_{Hall}$ became positive at $I_X$ = 0.2 mA. This sign reversal of the $V_{Hall}$ revealed that both hole and electron carriers existed in our nano-device. The holes resided primarily near the films surface region with high oxygen content (p type $TiO_2$ semiconductor channel) while the electrons resided in the interior region with low oxygen content (n type $TiO_\delta$ semiconductor, Schottky quantum well and metallic $Ti_2O$ channels). As sign of the $V_{Hall}$ changed from negative to positive, electron current gradually became larger than the hole current because most of the increasing current was transported by 2DEG. As $I_X$ further increased from 0.2 mA, amplitude of the $V_X$ started to decrease while the $V_{Hall}$ still increased, indicating that mobility of the 2DEG increased. According to the two-carrier model, 2DEG concentration and mobility could be obtained by the numerical fitting of $V_{Hall}$ and the fitting results were given in the Fig. 3(e). As $I_X$ increased from 0.2 to 0.6 mA, 2DEG concentration increased from 0.7 to 3.2 $\times 10^{13}$ cm$^{-2}$ and mobility increased from 90 to 210 cm$^2$ V$^{-1}$ s$^{-1}$. With increasing $I_X$, more electrons were distributed in the Schottky quantum well and depth of the potential well increased, which lead to mobility of 2DEG increased. The longitude magnetoresistance (MR) curves of the nano-device at $T$=3 K with different $I_X$ were shown in Fig. 3(f). For the 0.1 mA curve, the resistance increased monotonously with increasing magnetic field, which could be attributed to the ordinary magnetoresistance effect. For the 0.3 and 0.6 mA curves, the resistance decreased firstly and then increased with increasing magnetic field. The negative magnetoresistance at low magnetic field was attributed to the weak localization effect, which was a typical feature of 2DEG-transport in experiments [21,22].

Finally, we investigated the influence of temperature on the electrical properties of 2DEG with $I_X$ fixed at 0.6 mA. Electron concentration of the n type semiconductor channel $n_2$ increased with the increasing temperature. As a result, at higher temperatures, more electrons would be scattered into the Schottky quantum well under the same $E_X$ and contributed to the 2DEG transport. As shown in Fig. 4(a), when the temperature increased from 3 to 16 K, 2DEG concentration $n_3$ increased from 3.2 to 30 $\times 10^{13}$ cm$^{-2}$ whereas the mobility decreased from 210 to 70 cm$^2$ V$^{-1}$ s$^{-1}$ due to the electron-phonon and electron-electron scattering in the potential well were enhanced. The longitude MR curves under different temperatures were shown in Fig. 4(b) and Fig. 4(c). The amplitude of negative magnetoresistance at low magnetic field decreased with the increasing temperature and weak localization effect completely disappeared when $T$=10 K. According to the weak localization condition $k_F l_e \gg 1$ [23], where $k_F$ are Fermi wavevector, it can be concluded that the mean free path of the 2DEG $l_e$ decreased with increasing temperature.

In summary, this work has demonstrated that the Schottky quantum well can be formed in the titanium oxide thin films with the degree of oxidation gradually changed. According to the asymmetry interface electron scattering effect, the mobility and concentration of 2DEG formed in this potential well can be controlled by the in-plane electric field. These results show that electron redistribution induced by the asymmetry interface scattering, which can be expected to be observed in many other multilayer thin films as well, plays an important role in electron transport related process. Besides, the tunable rapid electron transport in the Schottky quantum well gives rise to the nonlinear resistance devices, which has potential applications in microelectronic industry, especially for the selectors in brain-inspired computing.

# ACKNOWLEDGMENTS

This work was sponsored by Chinese Academy of Sciences (grant No. QYZDY-SSW-JSC020, XDB44000000 and XDB28000000). This work was also supported by National Key R&D Program of China (No. 2017YFB0405700), and Beijing Natural Science Foundation Key Program (Grant No. Z190007).

# AUTHOR INFORMATION

Corresponding Author

[†]E-mail: lirunze@semi.ac.cn

**Notes**

Authors declare no conflicts of interest.

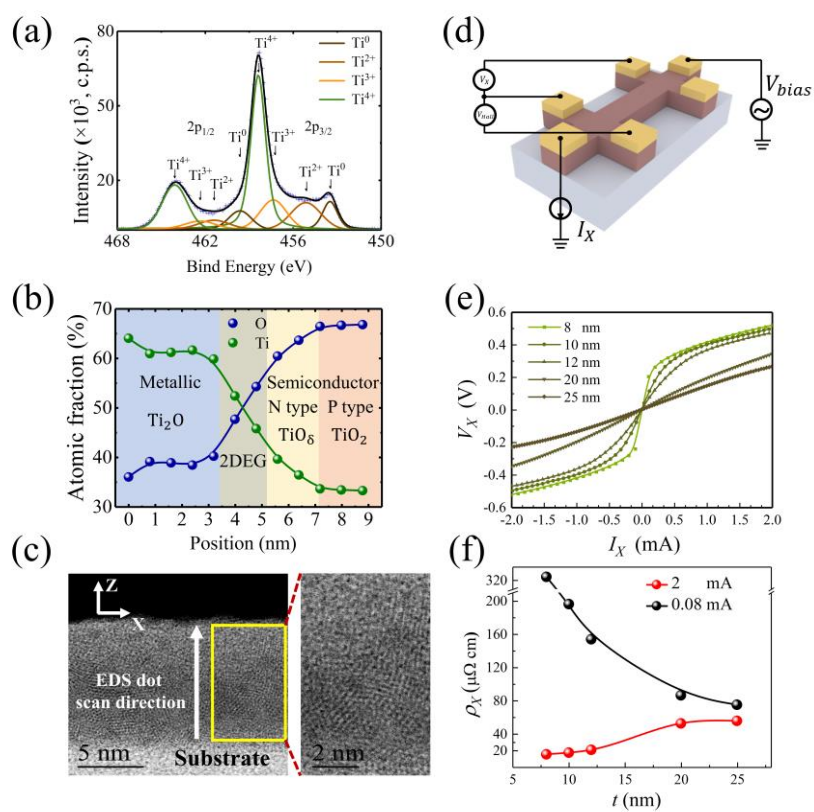

FIG. 1. Material characterization and electrical measurements at room temperature. (a) XPS spectrum of Titanium $2p_{3/2}$ and $2p_{1/2}$ for the naturally oxidized titanium film and peak fitting. (b) EDS dot-scan of atomic fraction of titanium and oxygen elements. The scan direction was shown by the white arrow in (c). (c) HR-TEM images of the 10 nm thick oxidized titanium film. (d) Schematic diagram of the six-probe electrical measurement technique and Hall device with 5 μm length and 1 μm width. (e) *VI* curves of Hall devices with different thicknesses. (f) shows the differential resistivity over the thickness with the $I_X$ fixed at 0.08 mA and 2 mA respectively.

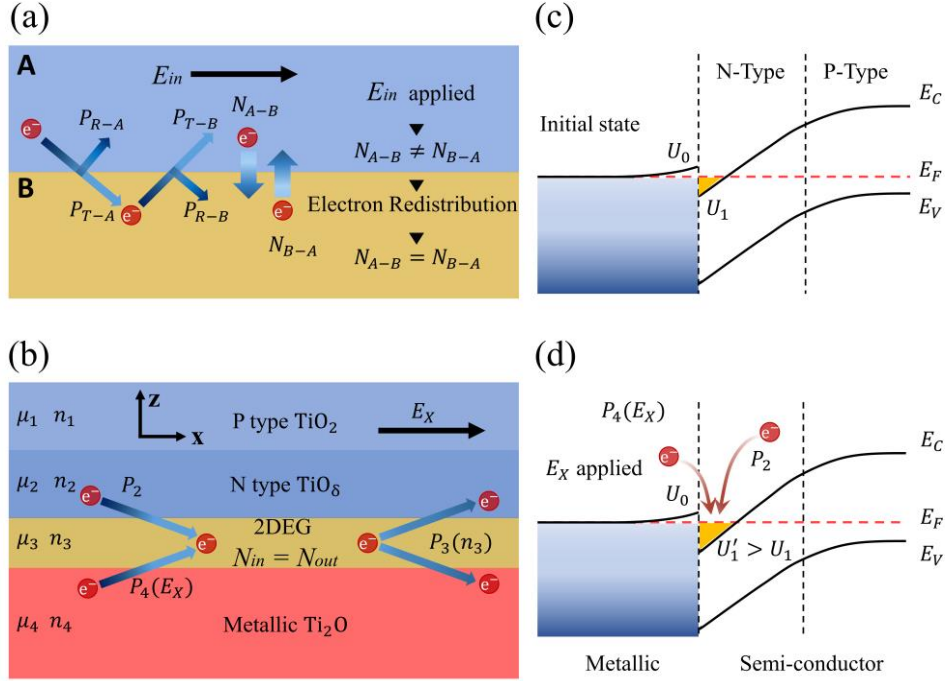

FIG. 2. Electron redistribution among different conductive channels. (a) Asymmetry interface electron scattering for a bilayer thin film. $P_{R\text{-}A, B}$ and $P_{T\text{-}A, B}$ are the reflection and transmission coefficient of A and B layers respectively. $N_{A\text{-}B}$ and $N_{B\text{-}A}$ are the interface scattered electron current density from A to B and B to A layer respectively. (b) The simplified physical conductivity model of our nano-devices with asymmetry interface electron scattering effect taken into consideration. (c) The band diagram with Schottky quantum well formed at the n type $TiO_\delta$ semiconductor/metallic $Ti_2O$ interface. $U_1$ is the depth of the formed Schottky quantum well. (d) Electrons were migrated from the n type $TiO_\delta$ semiconductor and metallic $Ti_2O$ channels to the Schottky quantum well channel with applied $E_X$ and thus the depth of the Schottky quantum well was increased.

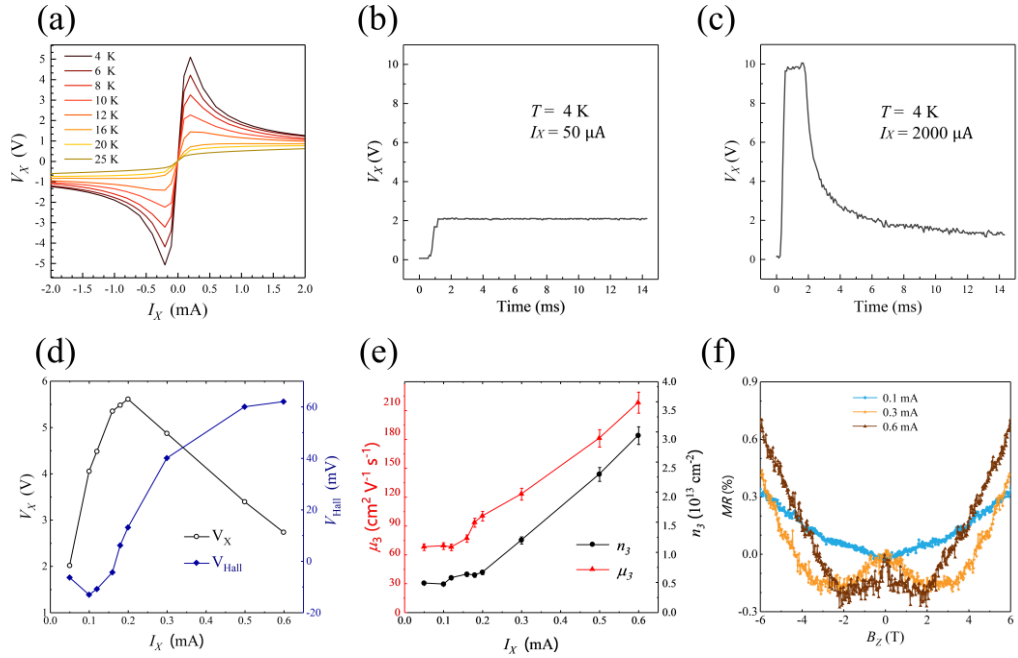

FIG. 3. Low temperature electron transport measurements for the 10 nm thick Hall device. (a) *VI* curves with different temperature. (b) and (c) are the changing processes of output voltage signal to reach the current level of 50 μA and 2000 μA respectively at $T$ =4 K. (d) $V_{Hall}$ (when $B_Z$ = +6 T) and $V_X$ changed with the current intensity at $T$ =3 K. (e) 2DEG concentration and mobility, which was obtained by the Hall voltage fitting results of (d), changed with the current intensity. (f) Longitude magnetoresistance curves with different applied $I_X$ when $T$ =3 K.

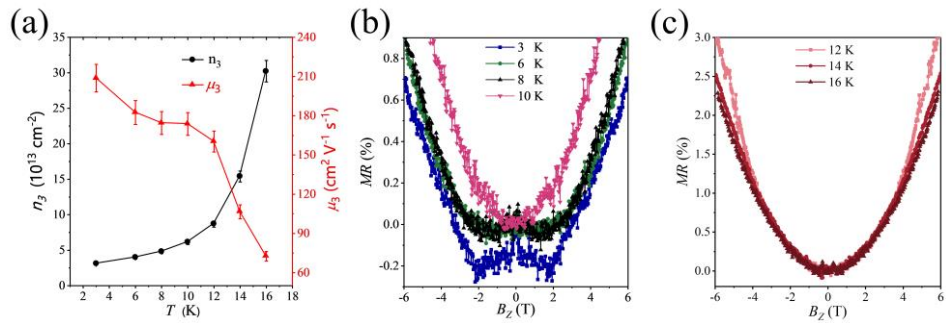

FIG. 4. Electron transport measurements for the 10 nm thick Hall device with $I_X$ fixed at 0.6 mA. (a) 2DEG concentration and mobility changed with the temperature. (b) and (c) Longitude magnetoresistance curves under different temperatures.